\documentclass[conference]{IEEEtran}
\usepackage{amssymb,amsmath,amsthm,amsfonts,mathtools}
\usepackage{graphicx}
\usepackage{amsmath}
\usepackage{color}
\usepackage{epstopdf}
\usepackage{cite}
\usepackage{bbm}
\usepackage{algorithm}
\usepackage{algpseudocode}
\usepackage[utf8]{inputenc}
\usepackage{cases}
\usepackage{gensymb}
\usepackage{subfloat} 
\usepackage{array}

\begin{document}

\title{\LARGE{Data-Driven Predictive Scheduling in Ultra-Reliable Low-Latency Industrial IoT: A Generative Adversarial Network Approach}}

\author{\IEEEauthorblockN{Chen-Feng Liu and Mehdi Bennis}
\IEEEauthorblockA{Centre for Wireless Communications, University of Oulu, Finland
\\E-mail: \{chen-feng.liu, mehdi.bennis\}@oulu.fi}
}

\maketitle

\begin{abstract}

To date, model-based reliable communication with low latency is of paramount importance for time-critical wireless control systems. In this work, we study the downlink (DL) controller-to-actuator scheduling problem in a wireless industrial network such that the outage probability is minimized. In contrast to the existing literature based on well-known stationary fading channel models, we assume an arbitrary and unknown channel fading model, which is available only via samples. To overcome the issue of limited data samples, we invoke  the generative adversarial network framework and propose an online data-driven approach to jointly schedule the DL transmissions and learn the channel distributions in an online manner. Numerical results show that the proposed approach can effectively learn any arbitrary channel distribution and further achieve the optimal performance by using the predicted outage probability.

\end{abstract}

\section{Introduction}
\label{Sec: Intro}

To enable ultra-reliable and low-latency communication (URLLC) \cite{Mehdi_URLLC}, a full characterization of wireless fading channels is crucial \cite{Swamy_thesis,LearningURLLC}, particularly, in industrial automation centered on stringent reliability and latency  \cite{5GACIA}.
Focusing on the uplink of an industrial Internet of things (IoT) setting, our previous work \cite{LiuIIoT} jointly studied the finite blocklength transmission and the tail distribution of the age of sensor's updated status information. Therein, we proposed a dynamic reliability and age-aware transmission policy for resource allocation and status updates, assuming instantaneous channel state information at the controller.
Taking into account the channel estimation error, Jurdi {\it et al.}~investigated the downlink (DL) outage probability in a multi-controller industrial network given  full information about channel fading and estimation noise \cite{Jurdi}. The vast majority of the existing literature, including the industrial IoT works \cite{LiuIIoT,Jurdi}, assume that channel fading is stationary within a coherence time. Some works further assumed that the parameters of the channel fading are available. In contrast, considering a mobile transmitter and receiver, 
Swamy {\it et al.}~showed that the channel fading varies within a coherence time\footnote{Thus, the outage probabilities of data retransmissions within a coherence time will not be identical.} and derived a closed-form expression of the fading channel correlation \cite{Swamy_thesis}. The correlation was further utilized to proactively minimize the transmission outage probability. 
The authors in \cite{LearningURLLC} considered the scenario in which the distribution family of the fading model is given but without the characteristic parameters. Since guaranteeing a certain reliability performance is challenging due to the imperfect channel model knowledge, the authors instead studied statistical reliability measures through the lens of  average reliability and probably correct reliability.

This work studies the DL scheduling problem of an industrial IoT scenario, where the controller needs to reliably send messages, e.g., control commands, to multiple actuators. Motivated by the works \cite{Swamy_thesis,LearningURLLC}, we take into account the fact that channel fading  varies within a coherence time, and further assume that the channel fading distribution is \emph{arbitrary} and \emph{unknown}. The objective is to minimize the outage probability of the scheduled DL transmissions. However, to calculate the outage probability, the information about the arbitrary and unknown channel distribution is needed.
To tackle this issue, we resort to generative adversarial networks (GANs) \cite{GAN} in machine learning, which provide a powerful tool to learn from data samples  and  approximate any arbitrary distribution, and propose an online data-driven approach for jointly scheduling the DL transmissions and learning the channel distribution.
The effectiveness of the proposed approach to approximate any arbitrary channel model is verified via simulations.

\section{System Model and Problem Formulation}

\begin{figure}[t]
\centering
	\includegraphics[width=0.8\columnwidth]{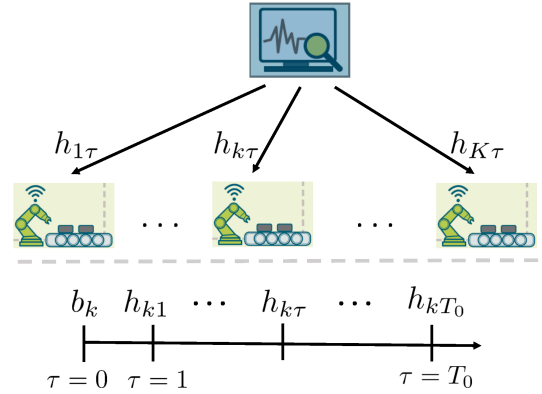}
	\caption{System model and time instants in a time slot.}
		\label{Fig: System}
		\vspace{-1em}
\end{figure}

\subsection{System Architecture}\label{Sec: System}

As shown in Fig.~\ref{Fig: System}, we consider the wireless industrial network which consists of a central controller and a set $\mathcal{K}$ of $K$ mobile robots/actuators. The goal is to schedule the DL, i.e., controller-to-actuator, transmissions in a time slot. Within a time slot, there are $T_0$ time instants denoted by $\mathcal{T}=\{1,2,\cdots,T_0\}$. The DL transmissions are scheduled and executed at the time instants. Specifically, all actuators' DL channel quality informations at the initial time instant, i.e., $\tau=0$, are available at the central controller. Then the controller schedules the DL transmission and sends the information to the actuator at a time instant $\tau\in\mathcal{T}$. We further consider that the length of the time slot is smaller than the coherence time length.
The scheduling indicator for each actuator $k\in\mathcal{K}$ in each instant $\tau\in\mathcal{T}$ is denoted by $ s_{k\tau}$ such that
\begin{subnumcases}{\label{Eq: Action set}}
\textstyle s_{k\tau}\in\{0,1\},&$\forall\,k\in\mathcal{K},\tau\in\mathcal{T},$\label{Eq: Scheduling indicator-1}
\\\textstyle\sum\limits_{\tau=1}^{T_k} s_{k\tau}= 1,&$\forall\,k\in\mathcal{K},$\label{Eq: Scheduling indicator-2}
\\\textstyle\sum\limits_{k\in\mathcal{K}}s_{k\tau}\leq S,&$\forall\, \tau\in\mathcal{T}.$\label{Eq: Scheduling indicator-3}
\end{subnumcases}
In \eqref{Eq: Scheduling indicator-1}, $ s_{k\tau}=1$ represents that  actuator $k$ is scheduled at time instant $\tau$. Otherwise, $ s_{k\tau}=0$. \eqref{Eq: Scheduling indicator-2} restricts that the transmission to actuator $k$ needs to be completed by the time instant $T_{k}$. We let $T_0=\underset{k\in\mathcal{K}}{\max}\{T_{k}\}$ for simplicity. In \eqref{Eq: Scheduling indicator-3}, the total number of simultaneous transmissions cannot exceed $S$. Additionally, we denote the set of all network-wide scheduling vectors $\mathbf{s}=[ s_{k\tau}:k\in\mathcal{K},\tau\in\mathcal{T}]$ which satisfy \eqref{Eq: Action set} as  $\mathcal{S}$ and assume that the total bandwidth is sufficient such that each scheduled actuator is dedicated an equal bandwidth.
As mentioned in Section \ref{Sec: Intro}, the fading channel coefficient varies over time owing to the actuator's mobility. In other words, the channel coefficient at the scheduled time instant $\tau\in\mathcal{T}$ will be  correlated with the coefficient measured at the initial instant $\tau=0$. Moreover, for the fading channel between the controller and actuator $k\in\mathcal{K}$, we denote the channel quality at $\tau=0$ as $b_{k}\in\mathcal{B}_k$ and the channel coefficient at each instant $\tau\in\mathcal{T}$ as $h_{k\tau}$, where $\mathcal{B}_k$ is a finite set. We further consider an arbitrary and unknown channel fading distribution.

\subsection{Problem Formulation}
\label{Sec: Problem}

Since the controller has only the realization of the channel quality vector $\mathbf{b}\!\!\!=\!\!\![b_{k}\!\!\!\!:\!\!k\!\!\in\!\!\mathcal{K}]$ when scheduling, we aim at minimizing the conditional outage probability, i.e., $\Pr\{\log_2\big(1+ S_{k\tau}\gamma|H_{k\tau}|^2\big) < \eta |\mathbf{B}=\mathbf{b}\}$, subject to the rate requirement $\eta$ and received signal-to-noise ratio (SNR) $\gamma$. To this end, we formulate the following problem as, $\forall\,\mathbf{b}$,
\begin{subequations}\label{Eq: problem}
\begin{IEEEeqnarray}{cl}
\hspace{-1.5em}\underset{\Pr(\mathbf{S}=\mathbf{s}|\mathbf{B}=\mathbf{b})\geq 0}{\mbox{maximize}}&\textstyle~~\sum\limits_{k\in\mathcal{K}}\sum\limits_{\tau\in\mathcal{T}}\Pr\big(|H_{k\tau}|^2 \geq \frac{2^{\eta}-1}{\gamma S_{k\tau}}\big|\mathbf{B}=\mathbf{b}\big)\label{Eq: problem-1}
\\\hspace{-1.5em}\mbox{subject to}&\textstyle~~\sum\limits_{\mathbf{s}\in\mathcal{S}}\Pr(\mathbf{S}=\mathbf{s}|\mathbf{B}=\mathbf{b})=1.
\end{IEEEeqnarray}
\end{subequations}
Here, the upper-case letters $\mathbf{S}$, $S_{k\tau}$, $\mathbf{B}$, $B_{k}$, and $H_{k\tau}$ represent random variables/vectors while the lower-case letters $\mathbf{s}$, $s_{k\tau}$, $\mathbf{b}$, $b_k$, and $h_{k\tau}$ represent the corresponding realizations. The goal in \eqref{Eq: problem} is to find the optimal probabilistic scheduling policy, which is challenging due to the lack of  conditional probability distribution function (PDF) of channel fading, i.e., $f(h_{k\tau}|B_k=b_k)$. Let us rewrite the objective function \eqref{Eq: problem-1} as
\begin{multline}
%
\textstyle\sum\limits_{\mathbf{s}\in\mathcal{S}}\sum\limits_{k\in\mathcal{K}}\sum\limits_{\tau\in\mathcal{T}}
\mathbb{E}_{h_{k\tau}}\big[\mathbbm{1}_{\big\{|H_{k\tau}|^2 \geq \frac{2^{\eta}-1}{\gamma s_{k\tau}}\big\}}\big|\mathbf{S}=\mathbf{s},\mathbf{B}=\mathbf{b}\big]
\\\textstyle\qquad\times\Pr(\mathbf{S}=\mathbf{s}|\mathbf{B}=\mathbf{b}).\label{Eq: Obj-1}
\end{multline}
From \eqref{Eq: Obj-1}, we can see that by empirically calculating the number of successful transmissions, i.e., $|h_{k\tau}|^2 \geq \frac{2^{\eta}-1}{\gamma}$, the expectation can be approximately found without the knowledge about $f(h_{k\tau}|B_k=b_k)$. However, since we focus on the URLLC regime in which the outage probability ranges from $10^{-9}$ to $10^{-5}$, the number of empirical transmissions to ensure sufficient times of failures, i.e., $|h_{k\tau}|^2 < \frac{2^{\eta}-1}{\gamma}$, will be tremendous.
To alleviate this shortcoming, we resort to the GAN which is a data-augmentation technique to enable us to synthetically learn any arbitrary distribution using historical channel realizations. 
With this in mind, we rewrite \eqref{Eq: Obj-1} as $\sum_{\mathbf{s}\in\mathcal{S}}\hat{\alpha}_{\mathbf{b}}(\mathbf{s})\Pr(\mathbf{S}=\mathbf{s}|\mathbf{B}=\mathbf{b})$ with
\begin{align}
%
\hat{\alpha}_{\mathbf{b}}(\mathbf{s})&\textstyle=\sum\limits_{k\in\mathcal{K}}\sum\limits_{\tau\in\mathcal{T}}
\int_{|h_{k\tau}|^2 \geq \frac{2^{\eta}-1}{\gamma s_{k\tau}}}\hat{f}(h_{k\tau}|\mathbf{S}=\mathbf{s},\mathbf{B}=\mathbf{b}){\rm d}h_{k\tau}\notag
\\&\textstyle=\sum\limits_{k\in\mathcal{K}}\sum\limits_{\tau\in\mathcal{T}}
\int_{|h_{k\tau}|^2 \geq \frac{2^{\eta}-1}{\gamma s_{k\tau}}}\hat{f}(h_{k\tau}|B_k=b_k){\rm d}h_{k\tau}
\label{Eq: Estimated PDF integral-1}
%
%
\end{align}
in which $\hat{f}(\cdot|\cdot)$ denotes the approximated conditional PDF. Here, we have $\hat{f}(h_{k\tau}|\mathbf{S}=\mathbf{s},\mathbf{B}=\mathbf{b})=\hat{f}(h_{k\tau}|B_k=b_k)$ since the channel coefficient $h_{k\tau}$ is independent of the scheduling vector $\mathbf{s}$ and the other actuators' channel quality $\mathbf{b}\setminus b_k$.
In the next section, we explain the steps of approximating $f(h_{k\tau}|B_k=b_k)$ and calculating \eqref{Eq: Estimated PDF integral-1} using GAN and detail our proposed online scheduling policy and the GAN-training approach.

\section{Online Data-Driven Approach for Joint Actuator Scheduling and GAN Training}
\label{Sec: Approach}

We first introduce the timeline of the online scheduling and training approach in which scheduling is done in a short timescale, whereas GAN training is executed over longer timescale. Specifically, the timeline is decomposed into frames indexed by $n\in\mathbb{Z}^{+}$, and each frame is composed of $M$ time slots (i.e., the time slot $\mathcal{T}$ in Section \ref{Sec: System}) indexed by $m\in\mathbb{Z}^{+}$. 
At the beginning of the $n$th time frame, a scheduling policy $\{{\pi}_{\mathbf{b}}(n;\mathbf{s})\!\!:\mathbf{b}\in\mathcal{B},\mathbf{s}\in\mathcal{S}\}$ is available at the controller. 
Then during the time frame $n$, observing a realization $\mathbf{b}^m$ at the beginning of each time slot $m\in[(n-1)M+1,nM]$, the controller selects a scheduling vector $\mathbf{s}^m$ based on the policy $\{{\pi}_{\mathbf{b}^m}(n;\mathbf{s})\!\!\!:\mathbf{s}\in\mathcal{S}\}$.
After finishing all scheduled transmissions in the $m$th slot, the controller is implicitly informed about each actuator $k$'s channel coefficient $h_{k\tau}^m$  at the scheduled time instant $\tau$. 
At the end of the time frame, the controller uses the channel realizations collected over all past $n$ time frames to train the GAN. The trained GAN provide us the approximated conditional PDF $\hat{f}(n;h_{k\tau}|B_k=b_k)$ which yields the probability (denoted by $\hat{\alpha}_{\mathbf{b}}(n;\mathbf{s})$ with the frame index $n$) in \eqref{Eq: Estimated PDF integral-1}.

\subsection{Generative Adversarial Networks}
\label{Sec: GAN}

Let us briefly explain GAN. GAN is a competitive game between a (synthetic data) generator and a (data) discriminator. When the generator mimicks the real data (e.g., the actual channel coefficients) to fool the discriminator, the goal of the discriminator is to distinguish  real data from  fake data.
The generator and discriminator are mathematically represented by the functions $\mathbf{x}=g_{\boldsymbol{\phi}}(\mathbf{z})$ (parameterized by $\boldsymbol{\phi}$) and $y=d_{\boldsymbol{\theta}}(\mathbf{x})$ (parameterized by $\boldsymbol{\theta}$), respectively. Here, $\mathbf{z}$ is a noise vector from a predetermined probability distribution, the vector $\mathbf{x}$ has the same size as both the real and synthetic data, and $y\in[0,1]$ indicates the likelihood of the authenticity of the input data.
Moreover, functions $g_{\boldsymbol{\phi}}(\cdot)$ and $d_{\boldsymbol{\theta}}(\cdot)$ can be trained using multilayer perceptrons (MLPs) \cite{GAN}, where the parameters $\boldsymbol{\phi}$ and $\boldsymbol{\theta}$ are composed of the weights and biases.
The generator and discriminator play the following two-player minimax game \cite{GAN}:
\begin{equation}\label{Eq: GAN problem}
\underset{{\boldsymbol{\phi}}}{\min}~\underset{{\boldsymbol{\theta}}}{\max}~\textstyle\mathbb{E}_{\mathbf{X}}\big[\log\big(d_{\boldsymbol{\theta}}(\mathbf{X})\big)\big]+\mathbb{E}_{\mathbf{Z}}\big[\log \big(1-d_{\boldsymbol{\theta}}(g_{\boldsymbol{\phi}}(\mathbf{Z}))\big)\big]
\end{equation}
in which the random vectors $\mathbf{X}$ and $\mathbf{Z}$ denote the real data and the input noise of the generator function, respectively.
Given a specific generator function $g_{\tilde{\boldsymbol{\phi}}}$, the optimal discriminator function is $d_{\boldsymbol{\theta}^{*}|g_{\tilde{\boldsymbol{\phi}}}}(\mathbf{x})=\frac{\Pr(\mathbf{X}=\mathbf{x})}{\Pr(\mathbf{X}=\mathbf{x})+\Pr(g_{\tilde{\boldsymbol{\phi}}}(\mathbf{Z})=\mathbf{x})},\forall\,\mathbf{x}$.
Further, the global optimality of \eqref{Eq: GAN problem} is achieved by the generator function $g_{\boldsymbol{\phi}^{*}}$ which satisfies $\Pr(g_{\boldsymbol{\phi}^{*}}(\mathbf{Z})=\mathbf{x})=\Pr(\mathbf{X}=\mathbf{x}),\forall\,\mathbf{x}$. 
In other words, the optimal generator can replicate the distribution of the real data. In this situation, the optimal discriminator is unable to differentiate between the real and synthetic data due to $d_{\boldsymbol{\theta}^{*}}(\mathbf{x})=1/2,\forall\,\mathbf{x}$. Then by using a large number of realizations of $\mathbf{Z}$ in the generator function $g_{\boldsymbol{\phi}^{*}}$, we can numerically build the distribution function of the real data.
To obtain the optimal generator function, i.e., $\boldsymbol{\phi}^*$, we iteratively and alternatively update the discriminator's and generator's parameters via stochastic gradient  descent (SGD) with \cite{GAN}
\begin{subequations}
\begin{align}
-&\frac{1}{L}\textstyle\sum\limits_{l=1}^{L}\nabla_{\boldsymbol{\theta}}\big[\log\big(d_{\boldsymbol{\theta}}(\mathbf{x}_l)\big)+\log \big(1-d_{\boldsymbol{\theta}}(g_{\boldsymbol{\phi}}(\mathbf{z}_l))\big)\big],\label{Eq: GAN_SGD-1}
\\&\frac{1}{L}\textstyle\sum\limits_{l=1}^{L}\nabla_{\boldsymbol{\phi}}\log\big(1-d_{\boldsymbol{\theta}}(g_{\boldsymbol{\phi}}(\mathbf{z}_l))\big),\label{Eq: GAN_SGD-2}
\end{align}
\end{subequations}
where $\mathbf{x}_{l}$ is one real data realization, $\mathbf{z}_l$ is one realization of the random noise vector, and $L$ is the size of a mini-batch. Note that in stead of \eqref{Eq: GAN_SGD-2}, we can consider the stochastic gradient $-\frac{1}{L}\sum_{l=1}^{L}\nabla_{\boldsymbol{\phi}}\log\big(d_{\boldsymbol{\theta}}(g_{\boldsymbol{\phi}}(\mathbf{z}_l))\big)$ for the generator's parameters to improve training performance \cite{GAN}.
%
The steps of training the GAN are detailed in Algorithm \ref{Alg: GAN}.
After the training completion, we obtain the conditional PDF $\hat{f}(n;h_{k\tau}|B_k=b_k)$ and $\hat{\alpha}_{\mathbf{b}}(n;\mathbf{s})$.

\subsection{Dynamic Updates for the Scheduling Policy}
\label{Sec: Scheduling}

Based on the probabilities $\{\hat{\alpha}_{\mathbf{b}}(n;\mathbf{s}):\mathbf{b}\in\mathcal{B},\mathbf{s}\in\mathcal{S}\}$, the best scheduling policy, $\forall\,\mathbf{b}\in\mathcal{B}$, is $\Pr(\mathbf{S}=\mathbf{s}^{*}|\mathbf{B}=\mathbf{b})=1$ with $\mathbf{s}^{*}=\underset{\mathbf{s}\in\mathcal{S}}{\arg\max}\,\hat{\alpha}_{\mathbf{b}}(n;\mathbf{s})$, which in turn is affected by the accuracy of the GAN's approximated conditional PDF. The more the channel realizations for training, the more accurate the approximation is. However, if the controller uses the scheduling policy $\Pr(\mathbf{S}=\mathbf{s}^{*}|\mathbf{B}=\mathbf{b})=1$ in the next time frame $n+1$, the same time instant is allocated to the actuator in all time slots with $\mathbf{B}=\mathbf{b}$. Thus, when the controller trains the GAN at the end of the next time frame $n+1$, the accuracy of the approximated conditional PDF for the other time instants cannot be further improved since there is no new training data. To address this concern, we instead consider 
\begin{subfigures}
\begin{figure}[t]
\vspace{-0.8em}
\end{figure}
\begin{algorithm}[t]
  \caption{GAN Training to Approximate the Arbitrary and Unknown Channel Distribution}
  \begin{algorithmic}[1]
  		\Require
		     $\mathcal{A}_{k\tau}^{n}(b_k)$, $E$ epochs, $C=5$, $L=20$, and Adam optimizer's parameters $(\psi,\beta_1,\beta_2)=(0.003,0.9,0.999)$.  
		\Ensure
		$\hat{\mathcal{A}}_{k\tau}^{n}$.
		\State Initialize $\boldsymbol{\theta}$ and $\boldsymbol{\phi}$.
		  \For{$e=1,\cdots,E$}
      \For{$j=1,\cdots,\big\lfloor \frac{|\mathcal{A}_{k\tau}^{n}|}{C\cdot L}\big\rfloor$}
     \For{$i=1,\cdots,C$}
  \State $\boldsymbol{\theta}\leftarrow \mbox{Adam}\big(-\frac{1}{L}\sum_{l=1}^{L}\nabla_{\boldsymbol{\theta}}\big[\log\big(d_{\boldsymbol{\theta}}(\mathbf{x}_l)\big)+\log \big(1-d_{\boldsymbol{\theta}}(g_{\boldsymbol{\phi}}(\mathbf{z}_l))\big)\big],\psi,\beta_1,\beta_2\big)$ with   $\{\mathbf{x}_l\}=\{a_l:[(j-1)C+i-1]L+1\leq l\leq [(j-1)C+i]L\}\subset\mathcal{A}_{k\tau}^{n}(b_k)$.
    \EndFor
\State $\boldsymbol{\phi}\leftarrow \mbox{Adam}\big(-\frac{1}{L}\sum_{l'=1}^{L}\nabla_{\boldsymbol{\phi}}\log\big(d_{\boldsymbol{\theta}}(g_{\boldsymbol{\phi}}(\mathbf{z}_{l'}))\big),\psi,$ $\beta_1,\beta_2\big)$.
     \EndFor
      \EndFor
        \State $\hat{\mathcal{A}}_{k\tau}^{n}(b_k)=\{g_{\boldsymbol{\phi}}(\mathbf{z}_{\tilde{l}}):1\leq \tilde{l}\leq 10^7\}$.
  \end{algorithmic}
\label{Alg: GAN}
\end{algorithm}
\end{subfigures}
\begin{multline}
\beta(\hat{\boldsymbol{\alpha}}_{\mathbf{b}}(n);\mathbf{s})\coloneqq\underset{\Pr(\mathbf{S}=\mathbf{s}|\mathbf{B}=\mathbf{b})}{\arg\max}\textstyle\sum\limits_{\mathbf{s}\in\mathcal{S}}\Pr(\mathbf{S}=\mathbf{s}|\mathbf{B}=\mathbf{b})
\\\times\big[\xi\cdot
\ln\big(\Pr(\mathbf{S}=\mathbf{s}|\mathbf{B}=\mathbf{b})\big)^{-1}+\hat{\alpha}_{\mathbf{b}}(n;\mathbf{s})\big]\label{Eq: tradeoff problem}
\end{multline}
with $\hat{\boldsymbol{\alpha}}_{\mathbf{b}}(n)=[\hat{\alpha}_{\mathbf{b}}(n;\mathbf{s}):\mathbf{s}\in\mathcal{S}]$ for notational simplicity. 
Note that $\xi(n)> 0$ is a time-variant parameter, which monotonically decreases with $n$, to trade off \emph{exploration} (i.e., maximizing information entropy) and \emph{exploitation} (i.e., maximizing the successful probability). 
When $n$ is small, the controller schedules the actuator at different time instants in successive time slots in order to have more channel realizations/training data in all time instants for each actuator. When the GAN is well trained as time elapses, the controllers will always schedule the actuator at the fixed time instant such that the sum of the approximated successful transmission probabilities is maximized.
The solution to problem \eqref{Eq: tradeoff problem} is
\begin{equation}\notag
\beta(\hat{\boldsymbol{\alpha}}_{\mathbf{b}}(n);\mathbf{s})=\frac{\exp\big( \hat{\alpha}_{\mathbf{b}}(n;\mathbf{s})/\xi\big)}{\sum_{\mathbf{s}\in\mathcal{S}}\exp\big( \hat{\alpha}_{\mathbf{b}}(n;\mathbf{s})/\xi\big)},~\forall\,\mathbf{s}\in\mathcal{S}.
\end{equation}
Finally, taking the cumulative moving average, the controller updates the scheduling policy  in a recursive manner as per
\begin{align}
&\hspace{-0.4em}{\pi}_{\mathbf{b}}(n+1;\mathbf{s})=\frac{1}{n}\textstyle\sum\limits_{\tilde{n}=1}^{n}\beta(\hat{\boldsymbol{\alpha}}_{\mathbf{b}}(\tilde{n});\mathbf{s})={\pi}_{\mathbf{b}}(n;\mathbf{s})\notag
\\&\hspace{-0.4em}+
\frac{\big[\beta(\hat{\boldsymbol{\alpha}}_{\mathbf{b}}(n);\mathbf{s})-{\pi}_{\mathbf{b}}(n;\mathbf{s})\big]}{n},~\forall\,n\in\mathbb{Z}^{+},\mathbf{s}\in\mathcal{S},\mathbf{b}\in\mathcal{B}.\label{Eq: Learning update}
\end{align}
The steps of the online data-driven approach are outlined in Algorithm \ref{Alg: Main mechanism}.
\begin{subfigures}
\begin{figure}[t]
\vspace{-0.8em}
\end{figure}
\begin{algorithm}[t]
  \caption{Online Data-Driven Approach for Joint Actuator Scheduling and GAN Training}
  \begin{algorithmic}[1]
    \State Initialize $n=1$ and $\mathcal{A}_{k\tau}^{0}(b_k)=\emptyset,\forall\,k\in\mathcal{K},\tau\in\mathcal{T},b_k\in\mathcal{B}_k$, and set an initial value for
    ${\pi}_{\mathbf{b}}(1;\mathbf{s}),\forall\,\mathbf{s}\in\mathcal{S},\mathbf{b}\in\mathcal{B}$.
  \Repeat
        \For{$m=(n-1)M+1,\cdots,nM$}
          \State Observing a realization $\mathbf{b}^m$, the controller makes a scheduling decision $\mathbf{s}^m$ based on $\{{\pi}_{\mathbf{b}^m}(n;\mathbf{s}):\mathbf{s}\in\mathcal{S}\}$.
        \EndFor
            \State  The controller collects the channel gains and updates  $\mathcal{A}_{k\tau}^{n}(b_k)\leftarrow \mathcal{A}_{k\tau}^{n-1}(b_k)\cup \{|h_{k\tau}^m|^2\big|b_{k}^m=b_k,s_{k\tau}^m=1,(n-1)M+1\leq m\leq nM\}$ $,\forall\,k\in\mathcal{K},\tau\in\mathcal{T},b_k\in\mathcal{B}_k$.
                   \State  The controller trains the GANs by following Algorithm \ref{Alg: GAN}.
                   \State By using $\hat{\mathcal{A}}_{k\tau}^{n}(b_k)$, the controller builds $\hat{f}(n;h_{k\tau}|B_k=b_k),\forall\,k\in\mathcal{K},\tau\in\mathcal{T},b_k\in\mathcal{B}_k$.
                   \State The controller finds $\hat{\alpha}_{\mathbf{b}}(n;\mathbf{s}),\forall\,\mathbf{s}\in\mathcal{S},\mathbf{b}\in\mathcal{B}$, and updates  \eqref{Eq: Learning update}.
               \State $n\leftarrow n+1$.
      \Until{Stopping criteria are satisfied.}
  \end{algorithmic}\label{Alg: Main mechanism}
\end{algorithm}
\end{subfigures}

\section{Numerical Results}
\label{Sec: Results}

We consider the 2.625\,GHz carrier frequency in a factory environment with one central controller and two actuators. Both actuators move at constant velocities of 5\,m/s and 10\,m/s and, hence, experience 22\,ms and 11\,ms coherence time. We assume that the time length between two instants is 1\,ms and $T_0=3$ such that the time slot length, i.e., 3\,ms, is shorter than the coherence time. In addition, $S=2$, $T_1=T_2=T_0$, and $M=5000$. Each scheduled sensor has a dedicated 10\,MHz bandwidth with a transmission duration of 1\,ms. The considered data sizes are 20 bytes and 250 bytes \cite{5GACIA}. Moreover, if the channel gain $|h_{k0}|^2$ at time instant $\tau=0$ is larger than 1, we denote the channel quality  as $B_k=1$. Otherwise, $B_k=0$. We simulate the fading channel model in \cite{Swamy_thesis}. The closed-form expression of the conditional successful probability \eqref{Eq: problem-1} based on this model is derived in the Appendix. Regarding GAN training, the generator's MLP consists of a 4-neuron input layer, a 8-neuron hidden layer, and  single-neuron output layer while the discriminator's MLP consists of a single-neuron input layer, a 24-neuron hidden layer, and single-neuron output layer.
The activation functions in the hidden layers of both the generator and discriminator are the leaky rectified linear unit (ReLU). For the  activation functions in the output layers, we consider $\tanh(\cdot)$ in the generator and the sigmoid function in the discriminator.
The input noise of the generator is based on a multivariate exponential distribution in which the composed random variables are independent and identically distributed with the marginal distribution  ${\rm Exp(1)}$. Moreover, we normalize the channel gains such that the input training data to the discriminator belong to $[-1,1]$. For performance comparison, we consider a baseline in which the controller is agnostic to the channel variation and schedules the actuators in a random manner. 

\setcounter{figure}{1}
\begin{figure}[t]
\centering
	\includegraphics[width=0.99\columnwidth]{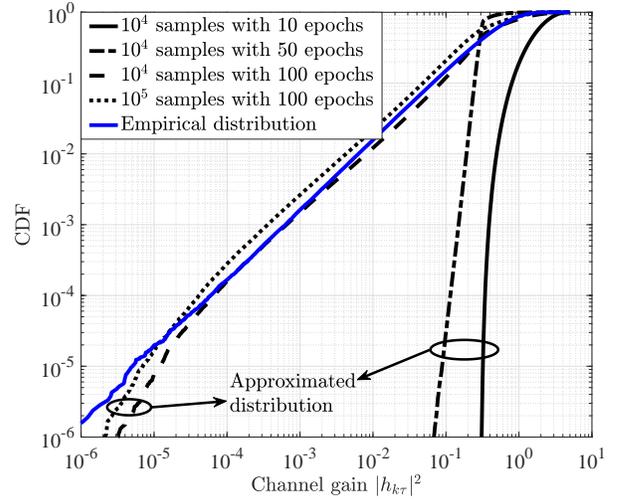}
	\caption{CDFs of the empirical distribution and approximated distributions for various training epochs and different amounts of training data/samples.  $B_k=0$, $k=2$, and $\tau=1$.}
		\label{Fig: 1}
		\vspace{-1em}
\end{figure}
\begin{figure}[t]
\centering
	\includegraphics[width=0.99\columnwidth]{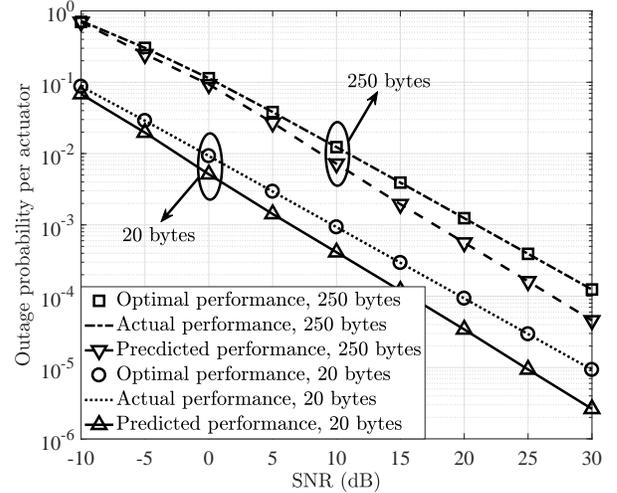}
		\caption{1) Outage probability of the optimal scheduling policy; 2) outage probability achieved by our proposed approach; and 3) predicted outage probability; versus SNR for different data sizes.}
		\label{Fig: 2}
		\vspace{-1em}
\end{figure}

Fig.~\ref{Fig: 1} plots the accuracy of the approximated arbitrary cumulative distribution functions (CDFs). 
As expected, the approximation becomes more accurate by training the GAN with more epochs.
Since there are more samples with very small values in the training process, the accuracy of the tail of the approximated distribution
 increases with the number of training samples.
In Fig.~\ref{Fig: 2}, we show the predicted outage performance based on the approximated channel distributions using GANs and the actual performance achieved by our scheduling policy.  
Due to the approximation error, there is a gap between the predicted performance and actual performance. Nonetheless, the controller is still able to make an optimal scheduling decision based on the approximated information. As shown in Fig.~\ref{Fig: 2}, our achieved performance and the optimal outage probability match very well.
Moreover, the prediction error gap increases as the outage probability decreases. 
This effect is caused by the higher error in the tail of the approximated distribution.
To further improve the prediction in this regime, we can incorporate results in extreme value theory which characterizes the tail of general probability distributions \cite{TComLiu}.

\begin{table}[t]
  \centering
  \caption{Conditional Outage Probability Given the Channel Quality at the Initial Time Instant. $k=2$}
  \renewcommand{\arraystretch}{1.25}
\begin{tabular}{|>{\centering}m{1cm}||>{\centering}m{1cm}|>{\centering}m{1cm}|>{\centering}m{1cm}|>{\centering\arraybackslash}m{1cm}|}
\hline
$\tau=0$&$\tau=1$&$\tau=2$&$\tau=3$&$\tau\to\infty$\\
\hline
\hline
$B_k=0$&$10^{-3.75}$&$10^{-3.81}$&$10^{-3.9}$&$10^{-3.95}$
\\\hline
$B_k=1$& $ 10^{-6.56}$ & $10^{-4.42}$& $10^{-4.05}$&$10^{-3.95}$
\\\hline
\end{tabular}
\label{Tab: 1}
\vspace{-1em}
\end {table}

\begin{figure}[t]
\centering
	\includegraphics[width=0.99\columnwidth]{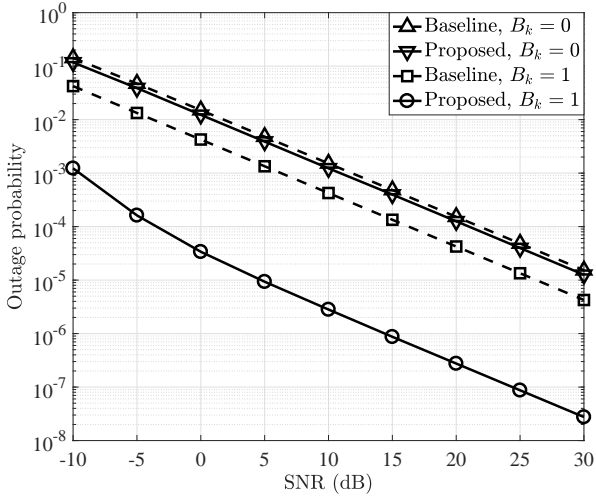}
	\caption{Outage probabilities of the proposed approach and baseline for both $B_k=0$ and $B_k=1$ as the SNR varies with the 20 bytes data. $k=2$.}
		\label{Fig: 3}
		\vspace{-1em}
\end{figure}

Finally, we compare the performances of our approach with the baseline.
Before showing the outage probability curves, let us emphasize the advantage of taking the channel correlation into account. Table \ref{Tab: 1} lists the conditional outage probability given the initial channel quality with $k=2$, the 20\,dB SNR, and the 20 bytes data. When the initial channel quality is bad/good, i.e., $B_k=0/1$, the conditional outage probability is high/low at the first time instant. As $\tau$ increases, the correlation diminishes such that the outage probability decreases/increases and converges. 
Therefore, by incorporating the channel correlation, the actuator will be scheduled at the nearest time instant if $B_k=1$. If $B_k=0$, the controller can schedule the actuator at the later time instant, i.e., a more uncorrelated channel fading realization. When $\tau\to\infty$, $B_k$ and channel fading $H_{k\tau}$ become independent. Therefore, the same converged conditional probability is achieved, irrespective of the value of $B_k$.
Since the channel correlation is considered in our scheduling approach, it outperforms the baseline in both $B_k=0$ and $B_k=1$ at various SNR values as shown in Fig.~\ref{Fig: 3}. Moreover, the performance superiority is more significant when $B_k=1$ because the channel correlation (which is reflected by the conditional outage probability in Table \ref{Tab: 1}) changes more rapidly in this regime.

\section{Conclusions}
\label{Sec: Conclusion}

In this work, we have studied the DL scheduling problem in an industrial IoT scenario in which the channel variation and correlation within a coherence time are taken into account.
We have further assumed that the channel fading model is arbitrary and unknown.
The lack of channel knowledge hinders us from solving the studied scheduling problem. 
To address this issue, we invoked the GAN framework to obtain the arbitrary distribution model by historical samples and further proposed an online data-driven approach to jointly schedule the actuators and train the GAN. Numerical results have shown the effectiveness of approximating the arbitrary and unknown distribution model.

\appendix
Given the channel coefficient  $h_{k0}$ at the initial time instant, we can find the conditional probability
\begin{multline*}
\textstyle\Pr\big(|H_{k\tau}|^2 \geq \frac{2^{\eta}-1}{\gamma s_{k\tau}}\big|H_{k0}=h_{k0}\big)
\\\textstyle= Q_1\Big(\frac{| J_0\big(\frac{2\pi v_k I\tau}{\lambda}\big)|\cdot |h_{k0}|}{\sigma_{k\tau}},\frac{1}{\sigma_{k\tau}}\sqrt{\frac{2^{\eta}-1}{\gamma s_{k\tau}}}\Big)
\end{multline*}
by referring to \cite{Swamy_thesis}. $v_k$ is the actuator $k$'s velocity, $I$ is the time length between two time instants, $\lambda$ is the carrier wavelength, $J_0(\cdot)$ is a first-kind Bessel function, $Q_1(\cdot,\cdot)$ is the Marcum Q-function, and $\sigma_{k\tau}=\sqrt{\frac{1}{2}-\frac{1}{2}\big[ J_0\big(\frac{2\pi v_k I\tau}{\lambda}\big)\big]^2}$. Then, incorporating the  channel gain  quantization interval  $[\underline{h}_{b_k},\bar{h}_{b_k}]$ for the channel quality value $b_k$, we can derive
\begin{multline*}
\textstyle\Pr\big(|H_{k\tau}|^2 \geq \frac{2^{\eta}-1}{\gamma s_{k\tau}}\big|B_k=b_k\big)
\\=\frac{\Pr(|H_{k\tau}|^2 \geq \frac{2^{\eta}-1}{\gamma s_{k\tau}},\underline{h}_{b_k}\leq |H_{k0}|^2\leq \bar{h}_{b_k})}{\Pr(\underline{h}_{b_k}\leq |H_{k0}|^2\leq \bar{h}_{b_k})}
%
%
%
\\=\frac{\int_{\underline{h}_{b_k}}^{\bar{h}_{b_k}}Q_1\Big(\frac{|J_0(\frac{2\pi v_k I\tau}{\lambda})|\sqrt{x}}{\sigma_{k\tau}},\frac{1}{\sigma_{k\tau}}\sqrt{\frac{2^{\eta}-1}{\gamma s_{k\tau}}}\Big) e^{-x}{\rm d}x }{e^{-\underline{h}_{b_k}}-e^{-\bar{h}_{b_k}}}.
\end{multline*}
%
%
%

\section*{Acknowledgments}
This research was supported by the Academy of Finland project CARMA, the Academy of Finland project MISSION, the Academy of Finland project SMARTER, and the Nokia Bell-Labs project ELLIS.

\bibliographystyle{IEEEtran}
\bibliography{Ref_WGAN}

\end{document}